\documentclass[intlimits,twoside,a4paper]{article}

\usepackage{amsmath,amssymb}
\usepackage{graphicx}
\usepackage{wrapfig}

\usepackage[T2A]{fontenc}
\usepackage[cp1251]{inputenc}

\usepackage{color}

\usepackage[]{cmpj2}

\articletype{Review Article}

\issue{2016}{19}{2}{22801}
\doinumber{10.5488/CMP.19.22801}

\title[Limits of Life:  Enzymes Under Pressure]%
{A molecular perspective on the limits of life: \\ Enzymes under pressure}
\author[Q.~Huang \textsl{et al}.]%
{Q. Huang\refaddr{label1}, K.N. Tran\refaddr{label1},
J.M. Rodgers\refaddr{label1,label2}, D.H.~Bartlett\refaddr{label3},
R.J.~Hemley\refaddr{label2}, T.~Ichiye\refaddr{label1}}
\addresses{
\addr{label1} Department of Chemistry, Georgetown University, Washington, DC 20057, USA
\addr{label2} Geophysical Laboratory, Carnegie Institution of Washington,
                Washington, DC 20015-1305, USA
\addr{label3} Scripps Institution of Oceanography, University of California, San
Diego, CA 92093-0202, USA}

\date{Received December 4, 2015, in final form January 14, 2016}
\authorcopyright{Q.~Huang, K.N.~Tran, J.M.~Rodgers, D.H.~Bartlett, R.J.~Hemley,
T.~Ichiye, 2016}

\begin{document}

\maketitle

\begin{abstract}
From a purely operational standpoint, the existence of microbes
that can grow under extreme conditions, or ``extremophiles'',
leads to the question of how the molecules making up these
microbes can maintain both their structure and function. While
microbes that live under extremes of temperature have been heavily
studied, those that live under extremes of pressure have been
neglected, in part due to the difficulty of collecting samples and
performing experiments under the ambient conditions of the
microbe. However, thermodynamic arguments imply that the effects
of pressure might lead to different organismal solutions than from the effects of temperature. 
Observationally, some of these solutions
might be in the condensed matter properties of the intracellular
milieu in addition to genetic modifications of the macromolecules
or repair mechanisms for the macromolecules. Here, the effects of
pressure on enzymes, which are proteins essential for the growth
and reproduction of an organism, and some adaptations against
these effects are reviewed and amplified by the results from molecular
dynamics simulations. The aim is to provide biological background
for soft matter studies of these systems under pressure.
\keywords enzymes, hydrostatic pressure, intracellular environment
\pacs 87.14.ej, 87.15.La, 87.15.B-
\end{abstract}

\section{Introduction}

The discoveries of ``extremophilic'' microbes and even higher
organisms that thrive under extremes of many conditions such as
temperature, pressure, salinity, pH, etc., raise many questions on
how life can exist under such conditions and what the limiting
conditions are, i.e., the ``limits of life''
\cite{Ichiye-ref01,Ichiye-ref02}. Fundamentally, determining the
adaptations for extreme conditions leads to a greater
understanding of all life at a molecular level. In addition,
understanding these adaptations can guide the search for life in
extreme environments such as beneath the continental and oceanic
surface or even extraterrestrially. Practically, understanding
these adaptations is also important for methods for sterilization
and food preservation by extreme conditions, which play critical
roles in human health and welfare. In addition, extremophilic
microbes and macromolecules from these organisms can play roles in
biotechnology, such as \textit{Thermophilus aquaticus} DNA
polymerase in polymerase chain reaction (PCR) techniques
\cite{Ichiye-ref03}. While there are clearly many types of
adaptations, one basic question is how the macromolecules making
up extremophiles can maintain their functional structure under
conditions that would destroy their counterparts in mesophiles
(moderate-loving).

Among the best-understood extremophiles are thermophiles
(hot-loving) and psychrophiles (cold-loving). Studies indicate
that these extremophiles employ both ``molecule-specific'' and
``global'' adaptive strategies to protect their macromolecules
against extremes of temperature. ``Molecule specific'' adaptive
strategies involve utilizing variations of molecules found in
mesophiles so that the molecules themselves are adapted for the
extreme, such as changes in the lipid composition of membranes
\cite{Ichiye-ref04,Ichiye-ref05} or genetic modifications of the
amino acid sequence of proteins \cite{Ichiye-ref06}. By contrast,
``global'' adaptive mechanisms protect general classes of
molecules in the organism against the extreme. These include heat-shock
and cold-shock proteins, many of which apparently function by
assisting in folding new proteins or refolding the damaged proteins.
Additionally, cryoprotectors and antifreeze proteins
\cite{Ichiye-ref07,Ichiye-ref08} are global mechanisms that
protect against cold by making changes in the physical properties
of the intracellular environment. Since the temperature limits
where microbial communities have been found range from $-20$ to
$122^{\circ}$C \cite{Ichiye-ref02}, this type of protection can be considered as ways the microbe 
find to ``cheat'' two phase
transitions of water, freezing and boiling, beyond the simple
colligative properties of freezing point depression and boiling
point elevation.

While the effects of temperature are heavily studied, pressure is
an underappreciated physical and thermodynamic parameter that has
influenced the evolution and distribution of life
\cite{Ichiye-ref09,Ichiye-ref10,Ichiye-ref11}. High-pressure
environments are the largest part of the biosphere and include the
deep sea, the sub-seafloor and the continental subsurface
\cite{Ichiye-ref02,Ichiye-ref12}. This represents $\sim 10^{30}$
microbial cells, a large fraction of total organism numbers,
biomass, and evolutionary history
\cite{Ichiye-ref13,Ichiye-ref14}. Microbes that grow best under
pressures greater than atmospheric pressure are termed piezophiles
\cite{Ichiye-ref15}. From the wide spread phylogenetic
distribution of piezophiles, it is apparent that piezophilicity
has evolved multiple times. However, piezophiles are among the
least understood extremophiles, in part because of the difficulty
in collecting samples and performing experiments at high
pressures. In addition, while freezing and boiling of water are
everyday phenomena, the upper limit of pressure at which microbes
have been found correspond to pressures of $\sim 1.4$~kbar ($1~\text{bar}
= 0.1~\text{MPa} \approx 1$~atm) \cite{Ichiye-ref16,Ichiye-ref17},
which is near the ultimate compressive strength of bone
\cite{Ichiye-ref18}. This is far beyond everyday experience, which is why physical intuition fails.

So far, the most studied piezophiles are mainly from deep ocean
environments. However, since deep ocean environments are mostly
cold, it is difficult to distinguish adaptations for high pressure
versus low temperatures. Because of this, more studies on the
combined effects of temperature and pressure, as well as salinity,
on microbes such as has been done for four species of
\textit{Halomonas} \cite{Ichiye-ref19} are warranted. Molecule
specific strategies have been found for the lipid composition of
membranes \cite{Ichiye-ref02}, although there is a debate over
whether there is genetic adaptation of proteins to pressure
\cite{Ichiye-ref20}. In addition, a few piezophiles have been
shown to preferentially accumulate certain osmolytes in response
to pressure, namely $\beta$-hydroxybutarate \cite{Ichiye-ref21}
and glutamate \cite{Ichiye-ref22}. This indicates they might be
``piezolytes'' that protect against hydrostatic pressure much like
cryoprotectors protect against freezing by making changes in the
physical properties of the intracellular environment. In addition,
while the accumulation of osmolytes may indicate that
piezophilicity might be connected to resistance to osmotic
pressure, the preferential accumulation of only certain osmolytes
indicates that this may be too much of a simplification. Of
course, other global mechanisms are likely to be important as
well.

Understanding the pressure resistance of mesophilic pathogenic
microbes is also important. High-pressure preservation of food
(called pascalization analogous to pasteurization), which relies
on killing microbes using pressures of 6 to 8 kbar, is becoming
popular since it does not greatly affect the nutritional value,
taste, texture, or appearance and does not involve chemical
preservatives \cite{Ichiye-ref01}. In addition, high-pressure
treatments may become important in sterilization, especially with
the increase in drug-resistant bacteria. Disturbingly, a
pioneering study indicated that some mesophilic microbes are capable
of surviving pressures above 1 GPa (10 kbar) \cite{Ichiye-ref23}.
Although originally met with skepticism \cite{Ichiye-ref24},
`directed evolution' experiments have shown that while the maximum
survival temperature could only be extended a few degrees, the
maximum survival pressure was extended to the GPa range
\cite{Ichiye-ref25,Ichiye-ref26}, although whether it was due to
changes in gene expression or some other biochemical response, or
selection of a small collection of survivors is not clear
\cite{Ichiye-ref26}. While many strategies are likely to be
involved, a clue about a global mechanism microbes might use to
survive pressure comes from the observation that a mesophile has
been shown to accumulate sucrose and fructose at high pressures
\cite{Ichiye-ref27}, which may increase the intracellular
viscosity. In addition, the halophilic (salt-loving)
\textit{Halobacterium salinarum} NRC-1, which accumulates high
intracellular concentrations of $\sim4$~M KCl, normally lives at
atmospheric pressure, but has been shown to survive pressures up
to at least 4~kbar \cite{Ichiye-ref28}. This indicates
that vitrification of the intracellular environment may also play a
role since 4~M KCl in aqueous solution at 1~GPa is near freezing
even at 298~K \cite{Ichiye-ref29,Ichiye-ref30} and its viscosity
can be estimated as almost 2~mPa-s compared to 0.89~mPa-s for pure
water at 1 bar based on pressure-temperature data for pure water
\cite{Ichiye-ref31} and aqueous salt solutions
\cite{Ichiye-ref32,Ichiye-ref33}.

{\looseness=-1%
The above-mentioned studies of deep ocean
piezophiles and of pathogenic mesophiles point out that there are
actually two limits of life: (1) the limits of growth, or being capable
of thriving and reproducing under the extreme, and (2) the limits of survival (or
viability), or being capable of enduring the extreme and thriving again when
conditions become more hospitable. These limits are important in
determining the conditions for discovering new microbial communities
and for killing pathogenic microbes, respectively, and may also
involve different timescales. For instance, microbial communities
may take millennia to adapt to a specific habitat so that the
entire genome may be evolved for the habitat, including multiple
molecule-specific changes in each protein sequence involving
hydrogen bonds, hydrophobic interactions, void volumes, etc.,
although non-harmful traits or mechanisms may also be retained
from their original habitat. Conversely, molecule-specific changes
may be hard to evolve in every protein of a pathogenic mesophilic
microbe during the time frame of developing resistance, and are
less likely to be preserved from more ancient ``extreme''
conditions if they are deleterious for mesophilic conditions.
Instead, a microbe might adapt or resurrect specific parts of the
genome for global mechanisms to help preserve the entire proteome.
For pressure, $\sim 1.4$~kbar [\cite{Ichiye-ref34} and Bartlett
{et al}., unpublished results] is currently an upper limit
for growth based mainly on observations of piezophiles from the
cold deep ocean trenches, and $\sim 8$ to 9 kbar
\cite{Ichiye-ref35,Ichiye-ref36} is currently an upper limit for
survival based mainly on the observation of the pressures where even the hardiest pathogenic microbes are killed.

}

\begin{figure}[!b]
\centerline{%
\begin{tabular}{ccc}%
\includegraphics[width=7.8cm]{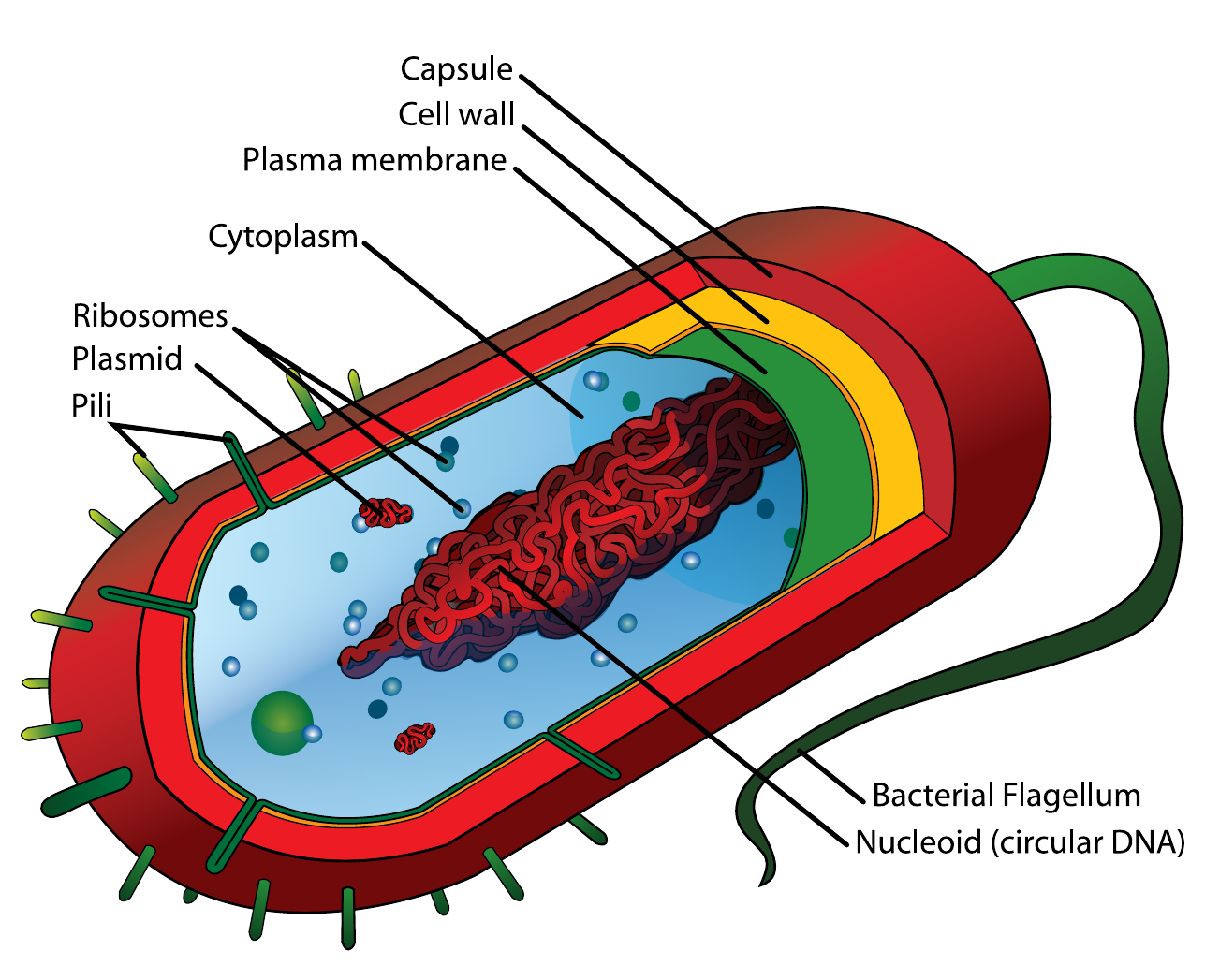} && \includegraphics{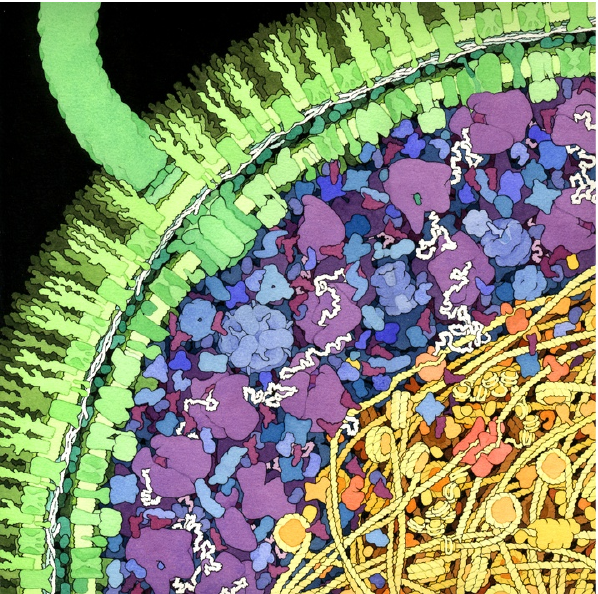} \\
(a) && (b)
\end{tabular}%
}%
\caption{(Color online) (a) Adapted from ``average prokaryote cell-en'' by
Mariana Ruiz Villarreal, Wikipedia. (b) A cross-section of a small
portion of an \textit{Escherichia coli} cell. The cytoplasmic area
is colored blue and purple. The large purple molecules are
ribosomes and the small, L-shaped maroon molecules are tRNA, and
the white strands are mRNA. Enzymes are shown in blue. The cell
wall is shown in green and the nucleoid region is shown in yellow
and orange. {Copyright David S. Goodsell, 1999.}}
\label{Ichiye-fig01}
\end{figure}

When pressure is applied to a microbe, the pressure is transmitted
into the intracellular domain [figure~\ref{Ichiye-fig01}~(a)].
Thus, the biomembranes and also the macromolecules inside the cell
feel the effects of pressure. While there are many complex
mechanisms involved in the overall growth and survival of a
microbe under pressure, there must also be a connection between
how macromolecules behave under pressure and how microbes live
under pressure. As mentioned above, microbes can alter the
chemical composition of the lipids in their membranes and also the
amino acid sequences of their proteins through evolution, but can
also respond more rapidly by changing the composition of the
intracellular co-solvent environment. However, the effects of
changes in the intracellular environment on macromolecules under
pressure are relatively unknown, and it is not even clear if they
are favorable for all types of biological macromolecules. For
instance, a piezolyte may favor the membrane fluidity but disfavor
the enzyme activity. While typical biochemical and biophysical studies
of biological macromolecules are carried out \textit{in vitro},
with perhaps salts and buffers added to the aqueous solution of
the protein, the intracellular environment \textit{in vivo} is a
complex concentrated mixture of other macromolecules, small
organic molecules, salts, and water
[figure~\ref{Ichiye-fig01}~(b)]. This leads to large differences
between the \textit{in vitro} and \textit{in vivo} environments,
including in the hydrogen bonding, hydrophobic effects, and
crowding, which are important for protein function.

At a molecular level, the range of conditions where macromolecules
will maintain activity by maintaining their \textit{functional}
structure can help define the limits for growth, while the range
where they will regain activity upon return to growth conditions
by maintaining their \textit{stable} but perhaps not functional
structure can help define the limits for survival. Since
protecting the functional structure against pressure has
requirements different from protecting stable structure, it is not clear if
either the molecule-specific or global protective strategies are
the same at the molecular level for both limits. In addition,
\textit{by understanding} \textit{the molecular mechanisms that
organisms use to withstand pressure, determining both limits could
be made more predictive rather than observational.}

The main focus here is on how enzyme activity, which is essential
for growth of an organism, responds to pressure. Interestingly,
the ``material'' requirements, or structural characteristics, for
enzymes may be opposite for growth under high pressure and
survival after high pressure. Both the structural requirements of
the protein for activity and how piezolytes could affect these
requirements are considered. Results from two sets of molecular
dynamics simulations using CHARMM36
\cite{Ichiye-ref49,Ichiye-ref50} at pressures between 1~bar and
10~kbar are used to illustrate certain points: one set is of 24~ns
simulations of GB1, the B1 domain of protein G, in TIP3P
\cite{Ichiye-ref81} water [Huang, Rodgers, and Ichiye, unpublished
results], and the other set is of 2~{\textmu}s simulations of
\textit{Clostridium acidurici} ferredoxin in TIP4P-Ew
\cite{Ichiye-ref51} water [Tran and Ichiye, unpublished results].

\section{Effects of pressure on proteins}

\begin{figure}[!b]
\centerline{
\includegraphics{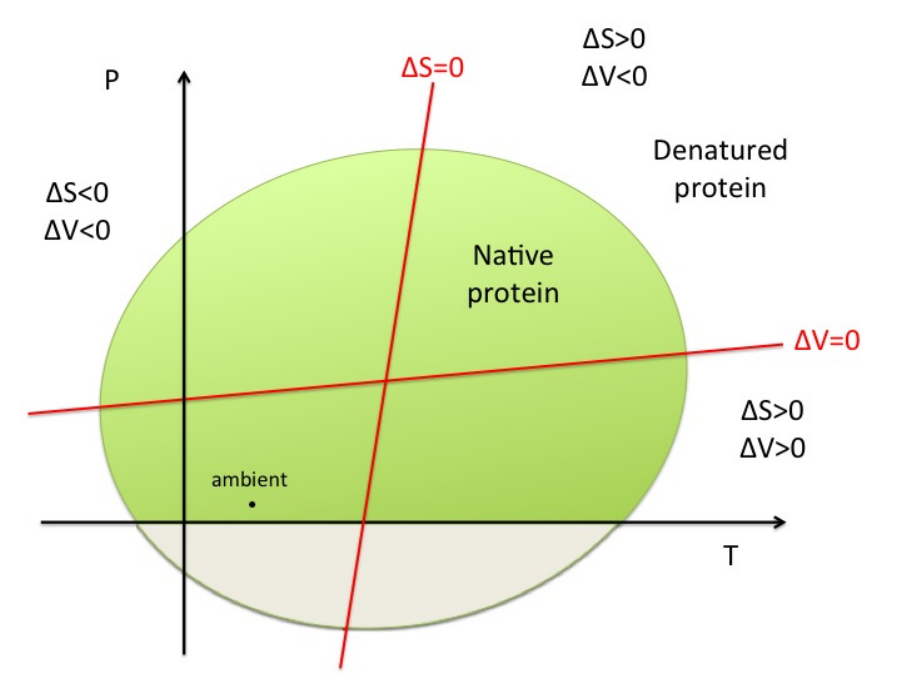}
}
\caption{(Color online) Schematic of a pressure-temperature stability diagram for
proteins.}
\label{Ichiye-fig02}
\end{figure}

Major effects of pressure on proteins are compression, making them
more compact and/or distorted, and unfolding
(figure~\ref{Ichiye-fig02}), as noted by Bridgman
\cite{Ichiye-ref37}. Since pressure-induced protein unfolding has
been studied extensively by many groups including Royer and
co-workers (i.e., reference \cite{Ichiye-ref39}), a brief but by no
means complete background is given here first, followed by a more
thorough discussion on the effects of pressure on proteins that are
relevant to enzyme activity.

\subsection{The limits of survival: protein unfolding and
oligomer dissociation}

From a molecular perspective, the disruption of protein structure to the
extent that normal structure cannot be regained in the intracellular milieu
upon the release of pressure may be a factor in the limits of survival. For
instance, complete unfolding of a protein would most likely lead to
non-specific aggregation within the cell upon release of pressure so that
refolding to the active state is not possible. In addition, since enzyme
activity often depends on being in the correct oligomeric state,
dissociation of oligomeric enzymes might also lead to non-specific
association with other molecules in the cytoplasm so that reassociation to
the functional oligomer is not possible.

Many studies of the sensitivity of proteins to pressure have
focused on the \textit{unfolding} of mesophilic proteins at high
pressures, which occurs between 4 to 8~kbar \textit{in vitro}
\cite{Ichiye-ref02}. By using pressure as another perturbant in
addition to temperature and chemical denaturants, important
insights can be gained in understanding the protein folding and
stability. To take advantage of high-pressure instrumentation,
studies have often utilized either mutants of staphylococcal
nuclease or T4 lysozyme, which unfold at unusually low pressures
without chemical denaturants. Although seemingly contrary to the
reduction of volume due to pressure, pressure unfolding is driven
by the reduction of volume of the entire system, which appears to
be due to changes in interactions between the polypeptide chain
and water. Although many specific effects have been proposed, it
appears to occur due to the loss of internal void volume in the
protein upon unfolding \cite{Ichiye-ref38}, or the ``destruction
of voids''.

\begin{figure}[!b]
\centerline{
\includegraphics{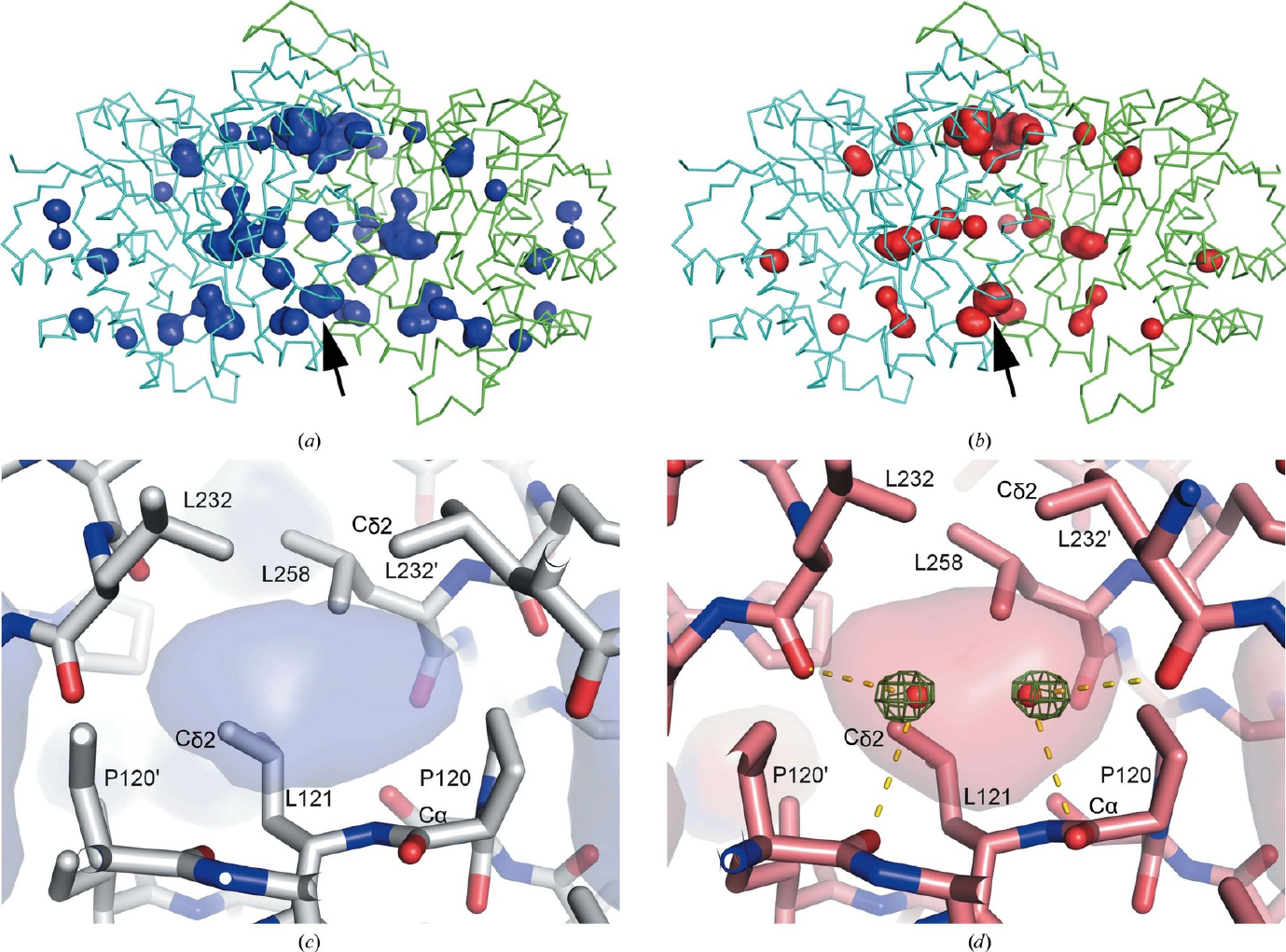}
}
\caption{(Color online) Internal cavities of the \textit{Shewanella oneidensis}
IPMDH dimer and observed water penetration. Internal cavities of
the dimer are shown as surface representations at (a) 1 and (b)
5.8~kbar; the cavity with the volume increase with increased
pressure is indicated by an arrow. In magnified views, water
inside this cavity, defined by transparent surfaces, is shown at
(c) 1 and (d) 5.8~kbar. Figures from Nagae~T., Ka\-wa\-mu\-ra~T.,
Chavas~L.M.G.,  Niwa~K.,  Hasegawa~M.,  Kato~C.,  Watanabe~N.,
{Acta Crystallogr. D}, 2012, \textbf{68}, 300.
Reproduced with permission of the International Union of Crystallography.
(\url{http://journals.iucr.org}).}
\label{Ichiye-fig03}
\end{figure}

Studies of mutants of staphylococcal nuclease have given a
compelling evidence that larger changes in internal volume due to
larger internal cavities lead to lower unfolding pressures
\cite{Ichiye-ref39}. In addition, complementary structural studies
using X-ray crystallography, NMR solution studies, and molecular
dynamics simulations have supported these results, although these
techniques tend to give more information on the folded state. For
instance, a crystallographic study of a T4 lysozyme L99A mutant at
increasing pressures showed up to four water molecules inside a
highly hydrophobic internal cavity created by the L99A mutation
starting at 1 to 2~kbar \cite{Ichiye-ref40}, suggestive of initial
stages in pressure-induced unfolding. However, the protein
remained folded up to 6.5~kbar even though fluorescence and
small-angle X-ray scattering studies indicate that the protein is
unfolded at these pressures \cite{Ichiye-ref41}. High-pressure NMR
studies of T4 lysozyme provided further support for the
``destruction of voids'' mechanism by showing that in the L99A
mutant, the domain with the mutation unfolds with increasing
pressure while a ``wild-type-like'' T4 lysozyme with no cavity and
a L99A mutant with benzene in the cavity do not unfold
\cite{Ichiye-ref42}. In addition, by comparing unconfined proteins
with proteins that were confined in reverse micelles that
prevented unfolding, these studies showed that the volume
reduction from pressured-induced unfolding in the unconfined
proteins was translated to increasing incorporation of water into
the cavity in the confined proteins, much like the
crystallographic experiments \cite{Ichiye-ref42}. This points to
the importance of crowding effects in the intracellular
environment [figure~\ref{Ichiye-fig01}~(b)].

Overall, compared to thermal or denaturant unfolding, two
important differences have been emerging: the pressure-induced
unfolded state appears to be more compact than the thermally
unfolded state \cite{Ichiye-ref43} and pressure unfolding appears
to involve extensive hydration in the interior of the protein
rather than exposure of the inner hydrophobic core to the bulk
solvent as in thermal unfolding \cite{Ichiye-ref44}. However, care
must be taken about interpreting the latter as a dynamic picture
of water being pushed inside the protein with increasing pressure,
since atomic fluctuations that would allow water to ``penetrate''
also decrease with increasing pressure. Instead, a better
interpretation may be a thermodynamic picture of a shifting
equilibrium of the populations of protein states towards states
with greater numbers of water molecules inside the cavities with
increasing pressure.

In addition, while unfolding of the entire protein by 8 kbar would
certainly limit the survival of a microbe, other less drastic
effects on proteins at lower pressures could also limit survival.
For instance, dissociation of oligomeric enzymes, which occurs
below 3~kbar \textit{in vitro} \cite{Ichiye-ref45}, will disrupt
their activity, so it may be a better determinant of the limits of
survival than complete unfolding. However, the intracellular
milieu may have two opposing effects on dissociation. In
particular, while reassociation of oligomers upon the release of
pressure would be made more difficult by non-specific association
in the heterogeneous environment, it might also be made easier
since the crowded intracellular environment might prevent
oligomers from completely dissociating. As in pressure induced
protein unfolding, water may play a role in oligomer dissociation
by pressure due to a ``destruction of voids'' mechanism. For
instance, crystallographic studies at different pressures of the
dimeric ($\sim 340$ residues/monomer) enzyme 3-isopropylmalate
dehydrogenase (IPMDH), a pressure sensitive enzyme in the
biosynthesis pathway of leucine, from the mesophilic
\textit{Shewanella oneidensis} \cite{Ichiye-ref46} shows water
inside a cavity at the interface of the two monomers between 4.1
to 5.8~kbar while no water is present at 1~bar
(figure~\ref{Ichiye-fig03}). This cavity may be a pressure
sensitive point for dimer dissociation.

\subsection{Limits of growth: compaction and conformational
changes of proteins}

From a molecular perspective, the perturbation of protein
structure to the extent that enzymes are no longer active should
be a factor in the limits of growth. For instance, the activity of
\textit{Escherichia coli} dihydrofolate reductase (DHFR) at 1~bar
was reduced to 65{\%} at 1~kbar \cite{Ichiye-ref47}, which
indicates that pressure can affect the enzyme activity. These perturbations
can be grouped into compaction and conformational changes.

\begin{figure}
\centerline{%
\begin{tabular}{ccc}%
\includegraphics{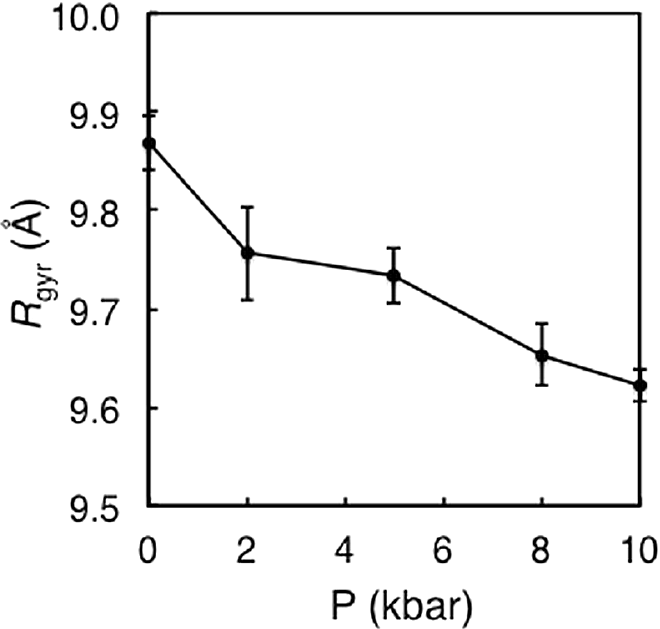} && \includegraphics{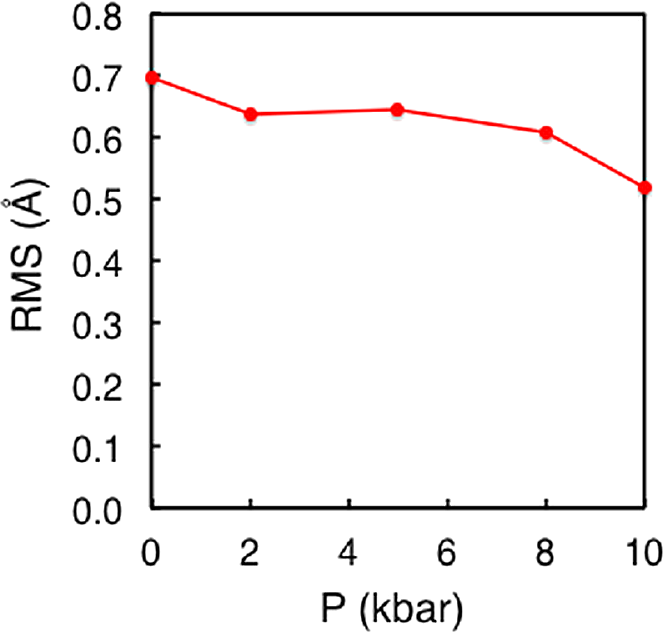} \\
(a) && (b)
\end{tabular}%
}%
\caption{(Color online) (a) Radius of gyration and (b) root mean-square
fluctuations of protein atoms as a function of pressure in
2~{\textmu}s molecular dynamics simulations of \textit{Clostridrium
acidurici} ferredoxin [Tran and Ichiye, unpublished results]. The
error bars for the radius of gyration correspond to standard
deviations to demonstrate the size of fluctuations.}
\label{Ichiye-fig04}
\end{figure}

The compaction of domains of protein has been demonstrated using
various structural methods. For instance, high-pressure NMR
solution studies of GB1, a small folding domain of protein G, show
that the domain compacts by $\sim 1{\%}$ between 30~bar and 2~kbar
\cite{Ichiye-ref48}. Compaction is illustrated by the changes in
the radius of gyration, $R_\text{gyr}$, of \textit{C. acidurici}
ferredoxin with pressure in 2~{\textmu}s molecular dynamics
simulations [Tran and Ichiye, unpublished results]
[figure~\ref{Ichiye-fig04}~(a)], which also show a $\sim 1{\%}$
compaction between 1~bar and 2~kbar. The results from these
simulations also indicate that {\textmu}s simulations are
needed to evaluate the changes in structural properties at different
pressures since very low frequency motions are apparent.
High-pressure crystallographic studies of monomeric and dimeric
proteins have been used to estimate the compressibility to be
between 4 to 6~Mbar$^{-1}$
\cite{Ichiye-ref46,Ichiye-ref52,Ichiye-ref53}, and internal
cavities within the monomers of IPMDH have been shown to be
compressed monotonously up to 6.5~kbar \cite{Ichiye-ref46},
indicating that the cavities allow the protein to be more compressible.

More important to enzyme function, compaction results in reduced
atomic fluctuations, which have often been noted as important for
enzyme activity. Reduction in atomic fluctuations with pressure is
illustrated by the root mean-square fluctuations of all protein
atoms in \textit{C. acidurici} ferredoxin from the 2~{\textmu}s
simulations [Tran and Ichiye, unpublished results]
[figure~\ref{Ichiye-fig04}~(b)], which show an average reduction
over the entire protein of about 10{\%} between 1~bar and 2~kbar.
Both the $R_{\mathrm{gyr}}$ and the atomic fluctuations show a
transition in behavior around 2 to 4~kbar. Additionally, an
analysis of the pressure dependence of fluctuations in
staphylococcal nuclease \cite{Ichiye-ref54} and lysozyme
\cite{Ichiye-ref55} using multiple short simulations shows a
similar transition around 4~kbar, which was attributed to the loss
of large amplitude, collective modes and restriction of
large-scale solvent translational modes. Since these large-scale
modes have often been implicated in the functional activity of
enzymes, the loss of such motions may be important in determining
the upper limit of growth.

In addition, compaction may lead to small perturbations of active
sites, including deformation since the compressibility of a
protein molecule is inhomogeneous, possibly leading to a decrease or
cessation of activity in an enzyme. In crystallographic studies of
yellow fluorescent protein (citrine) at pressures up to 5~kbar, a
shift in the fluorescence spectra between 1 and 2.8~kbar is
attributed to the progressive deformation of its chromophore by up
to 0.8~{\AA} \cite{Ichiye-ref56}. On the other hand, biochemical studies of
\textit{E.~coli} DHFR indicated that pressure did not affect the
hydride transfer, a chemical step, which indicates that no active site
distortion occurred \cite{Ichiye-ref57}. This suggests that it may
be very enzyme dependent how much distortion occurs with pressure
and how much will cause inactivation.

\begin{figure}
\centerline{
\includegraphics{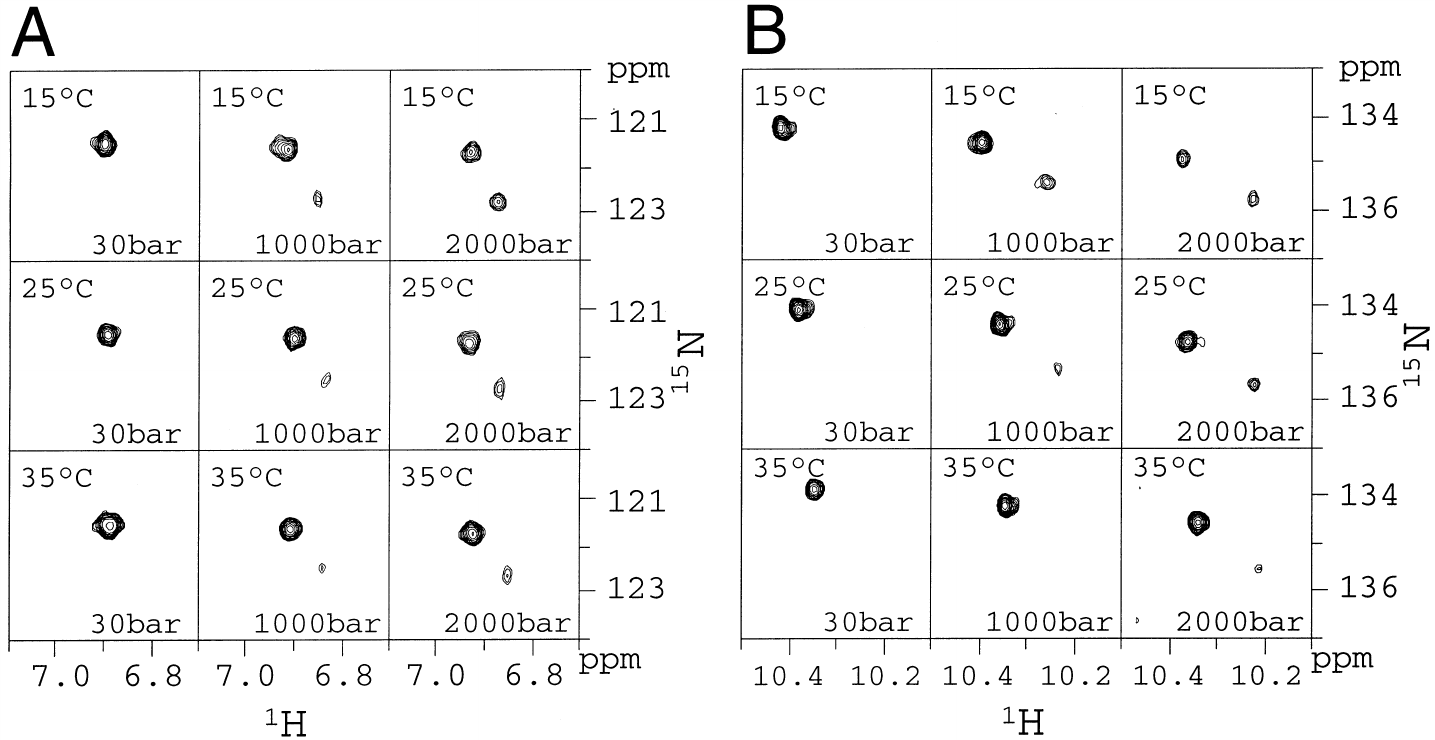}
}
\caption{The Val12-H and Trp22-N$\varepsilon $H cross-peaks in the
$^{15}$N/$^{1}$H HSQC spectra of \textit{Escherichia coli}
dihydrofolate reductase at various pressures and temperatures.
Reprinted with permission from Kitahara~R.,  Sareth~S.,  Yamada~H.,
Ohmae~E.,  Gekko~K.,  Akasaka~K., {Biochem.}, 2000, \textbf{39}, 12789.
Copyright American Chemical Society (2000).}
\label{Ichiye-fig05}
\end{figure}

Finally, pressure could induce conformational changes, or subtly,
shift populations of conformers that may play roles in different
stages of enzyme activity. While these could be the direct result
of compressive pressures stresses such as the loss of large amplitude modes
described above, the stable conformation may be determined by the
``destruction of voids'' mechanism. In fact, the observed changes
seem to fall in the latter category, since more open structures,
which would be easier to solvate, seem to be preferred at higher
pressures. For instance, the small monomeric ($\sim 200$ residues)
enzyme adenylate kinase (AK), which catalyzes the reversible
conversion between AMP/ATP and two ADP important to cellular
energy homeostasis, has large domain motion upon substrate binding
at atmospheric pressure based on X-ray crystallography
\cite{Ichiye-ref58}. \textit{E.~coli} DHFR shows large
conformational changes between 1.3 to 2.5~kbar in fluorescence
studies and although enzyme activity was not measured during
pressure treatment, it is presumed that the activity is destroyed by
conformational changes \cite{Ichiye-ref59}. Perhaps more
subtly, pressure can shift populations of conformations that
correspond to different steps of the catalytic mechanism. For
instance, DHFR has three conformations of the M20 (or Met20) loop
over the nicotinamide ring binding pocket based on X-ray
\cite{Ichiye-ref60} and NMR \cite{Ichiye-ref61} data that all
appear to play a role in its mechanism \cite{Ichiye-ref62}. While
the \textit{E.~coli} DHFR-THF binary complex is in the occluded
state at atmospheric pressure, high-pressure NMR studies show
increasing populations of the open state at pressures above 500~bar (figure~\ref{Ichiye-fig05}) \cite{Ichiye-ref47,Ichiye-ref62},
which may interfere with the reactive cycle. Also, high-pressure
NMR studies of ubiquitin indicate that pressure induces a transition
from a closed to an open conformation suitable for enzyme
recognition \cite{Ichiye-ref63}. Finally, IPMDH from the mesophile
\textit{S. oneidensis} has a sharp drop in activity above 50~kbar
\cite{Ichiye-ref64}, which appears to be due to the pressure
induced closure of the entrance to the active site that occurs
simultaneously with the opening of the groove of the active site
within the monomers of IPMDH \cite{Ichiye-ref65}.

\subsection{Protein stability under pressure}

The simulation results also illustrate why pressure induced
unfolding should be viewed as a thermodynamic rather than a dynamic
process. Analysis of the short 24~ns simulations of GB1 indicate
that water can readily enter and leave at 1 bar up to 2~kbar, but
above 2~kbar, does not enter if no water is initially present.
Even after 2~{\textmu}s, water does not start penetrating \textit{C.~acidurici} ferredoxin above 2~kbar up to 10~kbar even though the
latter is well above the unfolding pressures for typical proteins \textit{in vitro} 
(4 to 8~kbar) \cite{Ichiye-ref02}. This might seem contrary to the
above-mentioned observations of water inside cavities of a protein
only at high pressures in crystallographic studies of T4 lysozyme
\cite{Ichiye-ref40} and \textit{S. oneidensis} IPMDH
\cite{Ichiye-ref46} unless the thermodynamic and dynamic
viewpoints are considered simultaneously.

The thermodynamic picture is that the change in free energy for folding a
protein at constant $T$ and $P$ is
\begin{equation}
    \Delta G=\Delta H-T\Delta S=\Delta U+P\Delta V-T\Delta S.
\label{Ichiye-eqn01}
\end{equation}

Even though pressure is associated with forces, there is no
special force pushing water into the protein at high pressure just
as there is no hydrophobic force pushing the protein into an
unfolded state at high temperature. Instead, the protein molecules
with water inside them become more favorable as pressure is
increased according to equation~(\ref{Ichiye-eqn01}), since water
fills the voids and thus reduces the system volume in the unfolded
state.

However, water appears to find pathways into the protein in the
high-pressure crystal structures. At a molecular level, a simple
picture consistent with the simulation results and thermodynamics
is as follows. For a given cavity in the protein, a water molecule
can enter the cavity from bulk water by going over a barrier that
is lower than the barrier in a hypothetical rigid protein due to
the atomic fluctuations of the protein. At 1~atm, the atomic
fluctuations lower the barrier enough so that water can readily
enter and readily leave, as seen in the 20~ns GB1 simulations at 1
bar to 2~kbar. However, since the cavity has a low probability of
having water inside it thermodynamically
[equation~(\ref{Ichiye-eqn01})], the cavity is mostly empty at low
pressures. As pressure increases, the probability of water being
in the cavity increases thermodynamically
[equation~(\ref{Ichiye-eqn01})], since the overall volume of the
system is less when water is inside. However, the atomic
fluctuations of \textit{C.~acidurici} ferredoxin decrease with
pressure, as seen in the 2~{\textmu}s simulations, which
implies that the barriers for water penetration are larger. Thus,
while the equilibrium would be shifted toward water being inside
as pressure increases, it becomes a rare event for an individual
protein to reach that equilibrium state. Since the simulations of
\textit{C.~acidurici} ferredoxin begin with a single protein with
no water inside (the most favored state at~1 bar), the probability
of finding a water inside the protein at 10~kbar is very small
even after 2~{\textmu}s. Furthermore, the picture of empty cavities in
the protein waiting to be filled is simplistic in that some of the
cavities only appear due to atomic fluctuations, which are certainly reduced at high pressures. Altogether, the slow approach to
equilibrium indicates that very high pressures for a short
duration could be less disruptive to a protein than moderately
high pressures for a longer duration.

\subsection{Summary}

Of various effects of pressure on proteins, maintaining flexibility and
populations of loop conformations appear important in maintaining enzyme
activity under pressure. From a physical viewpoint, the effects of pressure
on general flexibility as measured by atomic fluctuations are a reflection
of the overall compressibility of a protein, a material science problem,
while the effects of pressure on populations of loop conformations may
involve the volume differences between the conformations.

\section{Protection against pressure}

While many mechanisms are most likely involved in protection
against pressure, the focus here is on the possible mechanisms
that could directly protect the enzyme activity in microbes by changes
in physical-chemical properties rather than on repair mechanisms. As
found in other extremophiles, a ``mol\-e\-cule-specific''
mechanism is that the sequence of the enzyme could make operation
of the enzyme more favorable under pressure. Additionally, a
``global'' mechanism is that the composition of the intracellular
environment could influence the enzyme activity under pressure.

\subsection{Molecule specific modifications of enzymes}

One type of a protective mechanism can be found by comparing
homologous proteins from ex\-trem\-o\-phil\-es and mesophiles,
which will identify evolutionary timescale ``molecule-specific''
mechanisms involving genetic mutations to protect each protein in
the proteome. For temperature, the ``flexibility-matching''
strategy has been noted by comparing homologous proteins from a
psychrophile (low-tem\-per\-a\-ture loving) and a thermophile
(high-temperature loving) with a mesophile \cite{Ichiye-ref66}. In
this strategy, the flexibility of the proteins from the
extremophiles at their growth temperature matches the flexibility
of the protein from the mesophile at standard temperatures and
pressures. In addition, comparisons of homologous proteins from
thermophiles and mesophiles indicate two different ways of
achieving flexibility matching, thermophilic archaea generally
have more hydrogen bonds while thermophilic bacteria generally
have a few more salt-bridges \cite{Ichiye-ref67}. Since archaea
are thought to have evolved first in high temperature environments
while bacteria are thought to have evolved at lower temperatures
but became adapted to some high temperature environments, the
observed adaptations are consistent \cite{Ichiye-ref67}. In
particular, it is easier to lose thermophilicity by losing
multiple hydrogen bonds than it is to gain it by adding multiple
hydrogen bonds, and it is easier to gain thermophilicity by adding
a few salt-bridges than by adding multiple hydrogen bonds.

Moreover, the need for flexibility matching has led to contrary
requirements for proteins from psychrophiles: while increasing
atomic fluctuations by weaker intramolecular interactions leads to
an increasing flexibility needed for functioning at low temperatures,
they also lead to less stable proteins that are more readily
unfolded. In fact, it has been noted that the cold-induced unfolding
temperature, $T_{\text{u}}$, for enzymes from psychrophiles is
actually higher for the homologous enzymes from mesophiles, which
has led to an activity-stability-flexibility hypothesis for enzyme
function for psychrophiles \cite{Ichiye-ref68}. In short, it
appears that maintaining flexibility is more important than
stability, as long as the enzyme is stable enough to maintain
sufficient structure.

Since an increasing pressure or a decreasing temperature can be
expected to reduce atomic fluctuations, this indicates that
piezophiles may also adapt by increasing the flexibility of their
proteins so that their fluctuations at high pressure are similar
to the fluctuations of proteins from mesophiles at atmospheric
pressure. In addition, an interesting correlation has been made
between cold and pressure unfolding of proteins
\cite{Ichiye-ref69}. However, since most piezophiles that have
been studied are from cold (but not freezing) deep ocean
environments, it is difficult to separate the effects of low
temperature and high pressure. Thus, there is a debate over whether
proteins are actually adapted to high pressure
\cite{Ichiye-ref20,Ichiye-ref70}. Intriguingly, there are also
piezophiles that are thermophilic \cite{Ichiye-ref02}, so more
studies of proteins from these organisms would be of great
interest.

\begin{figure}[!b]
\centerline{%
\begin{tabular}{cc}%
\includegraphics{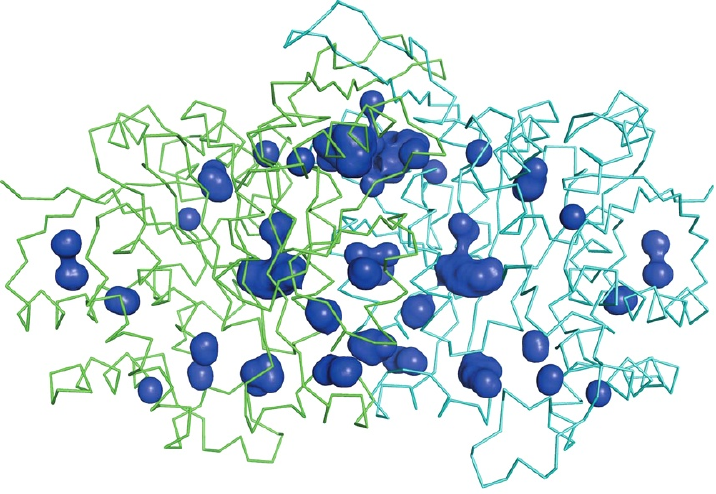} & \includegraphics{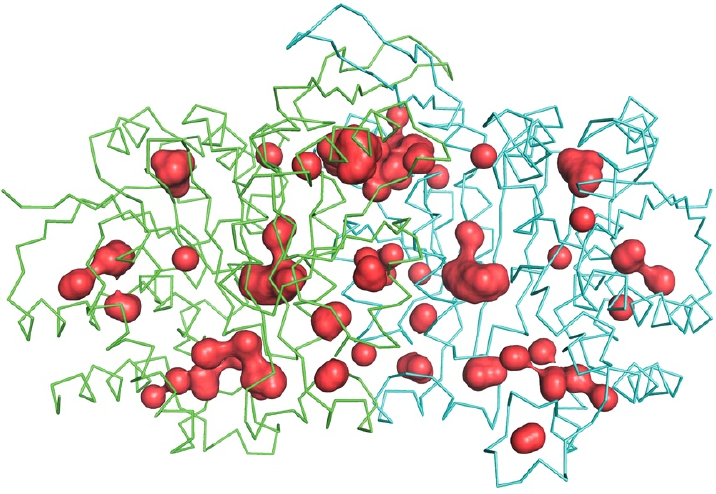} \\
(a) & (b)
\end{tabular}%
}%
\caption{(Color online) The internal cavities of (a) \textit{S.~oneidensis} IPMDH
(blue) and (b) \textit{S.~benthica} IPMDH (red) dimers. The wire
representation shows the overall structure of the IPMDHs (each
subunit is drawn in green and cyan). Figures from Nagae~T.,
Kato~C.,  Watanabe~N., {Acta Crystallogr. F}, 2012, \textbf{68}, 265.
Reproduced with permission of the International Union of
Crystallography. (\url{http://journals.iucr.org}).}
\label{Ichiye-fig06}
\end{figure}

The question regarding the relative balance between an increased
flexibility for activity over a decreased flexibility for stability
for enzymes from piezophiles can be examined by comparing
homologous enzymes from piezophiles and mesophiles.
Crystallographic studies of IPMDH from the obligate piezophile
\textit{Shewanella benthica}, with the growth pressures of 0.7 to
1~kbar \cite{Ichiye-ref71}, and the mesophile \textit{S.
oneidensis} have been performed \cite{Ichiye-ref65}. The
piezophile IPMDH has a more open structure with a larger internal
cavity volume than the mesophile IPMDH
(figure~\ref{Ichiye-fig06}), which would seemingly make it more
susceptible to pressure unfolding. Instead, a larger internal
cavity volume was proposed to make the protein more compressible
and less subject to pressure-induced distortion, thus allowing it
to remain active at higher pressures. In particular, since the
piezophile IPMDH retains almost the same $k_{\text{cat}}$ up to
2~kbar while that from the mesophile drops sharply above 50~bar
\cite{Ichiye-ref64} and there are no other significant differences
in the crystal structures, the larger void volume may help
maintain protein flexibility at a higher pressure. However, it
must be noted that \textit{S.~oneidensis} MR1 may not be a typical
mesophile since it was isolated from Lake Oneida in New York
State, which is a shallow freshwater lake that freezes over
completely in  winter \cite{Ichiye-ref72}. Thus, it is capable of growing
over a wide range of temperatures including near
$0^{\circ}$C so that the differences between \textit{S.~benthica} and
\textit{S.~oneidensis} may not be purely reflective of pressure
adaptation.

\begin{figure}
\centerline{%
\begin{tabular}{cc}%
\includegraphics{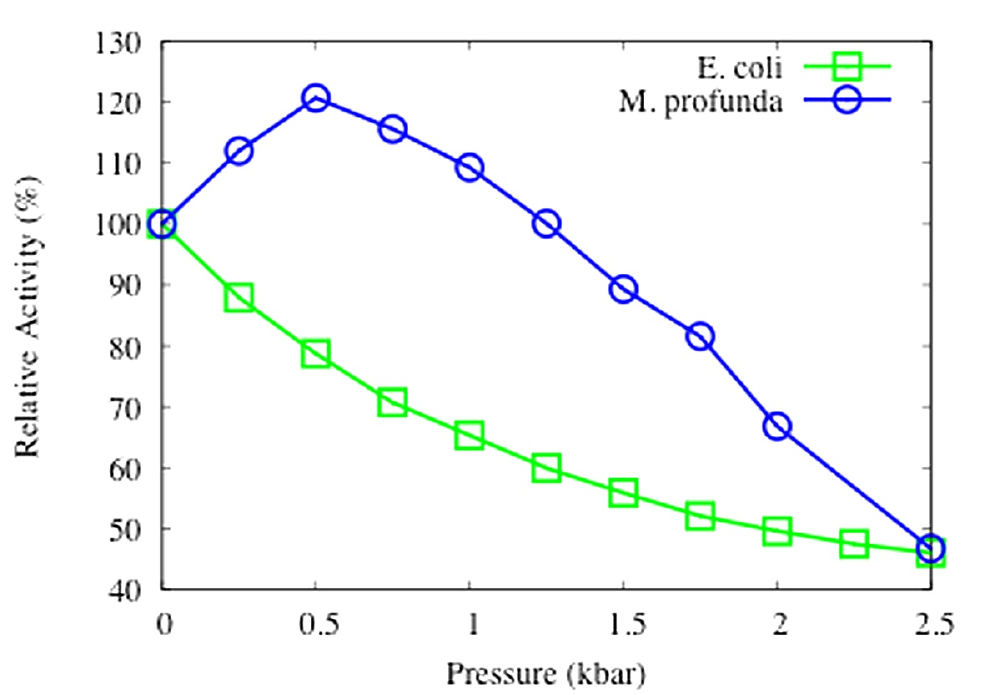} & \includegraphics{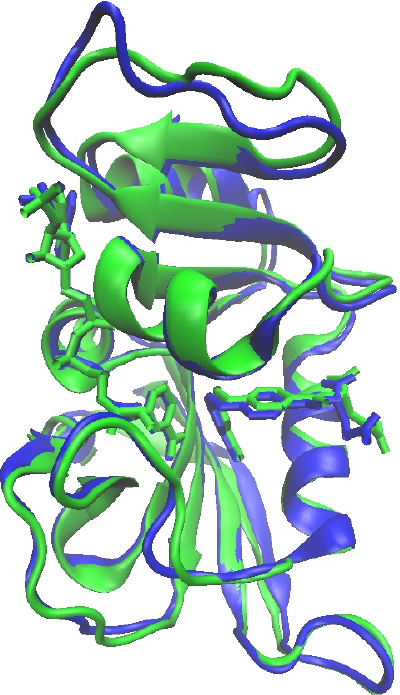} \\
(a) & (b)
\end{tabular}%
}%
\caption{(Color online) (a) Pressure dependence of relative enzyme activity of
\textit{E.~coli} DHFR (filled circle) and \textit{M.~profunda}
DHFR (open circle) at $25^{\circ}$C and $\text{pH} = 7$. Figure redrawn from
Ohmae~E.,  Murakami~C.,  Tate~S.-i., Gekko~K.,  Hata~K.,  Akasaka~K.,
 Kato~C., {Biochim. Biophys. Acta, Proteins Proteomics}, 2012, \textbf{1824}, 511.
Copyright with permission from Elsevier, 2012. (a)
Crystal structures of DHFR with NADP+ and folate from \textit{E.~coli} (green, PDB ID: 1RX2) and \textit{M.~profunda }(blue, PDB ID: 2ZZA).}
\label{Ichiye-fig07}
\end{figure}

In another comparison, the facultative psychropiezophile
\textit{M.~profunda} DHFR (55{\%} sequence identity with
\textit{E.~coli} DHRF) was found to display an increased activity
followed by a decreased activity as a function of an increasing
pressure, a feature found in some piezophiles
[figure~\ref{Ichiye-fig07}~(a)] \cite{Ichiye-ref74} but the cause is not apparent in
the crystal structures of \textit{E.~coli} and \textit{M.
profunda} DHFR [figure~\ref{Ichiye-fig07}~(b)] \cite{Ichiye-ref20}.
Since the solution conditions were the same for mesophile and
piezophile DHFR activity studies, a molecule-specific modification
could be responsible for the initial increased activity observed
in the piezophile. For instance, \textit{M.~profunda} DHFR is more
flexible than \textit{E.~coli} DHFR at atmospheric pressures
\cite{Ichiye-ref73}, indicating a flexibility matching mechanism
for protection. Specifically, \textit{M.~profunda} DHFR may have
a reduced activity at 1 atm because it is too flexible and reaches
both optimal flexibility and activity at $\sim 0.5$~kbar. In
addition, the subsequent decrease in the activity of \textit{M.~profunda} DHFR is most likely due to a similar mechanism
responsible for the decrease in \textit{E.~coli} DHFR. For
instance, the opening of the M20 loop with pressure noted in
\textit{E.~coli} DHFR may occur in \textit{M.~profunda} DHFR,
since the open state also has been seen in crystal structures of
\textit{M.~profunda} DHFR \cite{Ichiye-ref57}. The M20 loop
opening was noted above as a ``destruction of voids'' mechanism.
Interestingly, the unfolding pressure for \textit{M.~profunda}
DHFR (0.66 to 0.73~kbar between 15.6 and $28.8^{\circ}$C) is much
lower than for \textit{E.~coli} DHFR (2.58 to 2.72~kbar between
15.2 and $27.0^{\circ}$C) \cite{Ichiye-ref74}, indicating a
similar trend as the activity-stability-flexibility hypothesis for
psychrophiles. However, it should be noted that an activity
maximum at higher pressures is not always associated with
piezophiles, as noted in a study of six homologous DHFR from
different species of \textit{Shewanella} bacteria
\cite{Ichiye-ref75}. The lack of a maximum at high pressure for a
piezophile could reflect that the absolute (rather than relative)
activity is sufficient at its growth pressure, or that other
factors such as the intracellular environment enhance its
activity.

\subsection{Global changes in the intracellular milieu}

At least circumstantial evidence exists that accumulation of
certain co-solutes is a response to pressure. In particular,
\textit{$\beta $}-hydroxybutyrate monomers and oligomers
accumulate in the deep-sea bacterium \textit{P.~profundum} SS9,
which grows optimally at $15^{\circ}$C and 0.28~kbar
\cite{Ichiye-ref76} and glutamate accumulates in the hydrothermal
vent bacterium \textit{Desulfovibrio hydrothermalis} sp. nov.,
which grows optimally at $35^{\circ}$C and better at 0.26~kbar
than 1~bar \cite{Ichiye-ref77}. In addition, the mesophilic
\textit{Lactococcus lactis} has been shown to accumulate sucrose
and fructose at high pressures \cite{Ichiye-ref27} and \textit{H.~salinarum} NRC-1, which accumulates high intracellular
concentrations of K$^{+}$ and Cl$^{-}$ at similar molarity to
hypersaline environments ($\sim 4$~M NaCl) \cite{Ichiye-ref78},
normally lives at atmospheric pressure but can survive pressures
up to at least 4~kbar \cite{Ichiye-ref28}. However, it is not
clear that the protective mechanisms for piezophiles and microbes
that normally live at atmospheric pressure are the same at a
molecular level, or whether they should be. For instance,
an increased viscosity of the environment probably protects proteins
against pressure-induced unfolding in a mesophile to survive high
pressure while it is not clear whether it is more important to
protect the stability or the flexibility of proteins in a
piezophile. However, it may be possible  that piezophiles have
proteins with genetic modifications for flexibility but use
piezolytes to stabilize them against unfolding. In addition, it is
not even clear if the accumulation of these co-solutes is
protective of, deleterious to, or neutral for protein function, as
they probably protect other parts of the microbe.

\begin{wrapfigure}{I}{0.5\textwidth}
\centerline{
\includegraphics{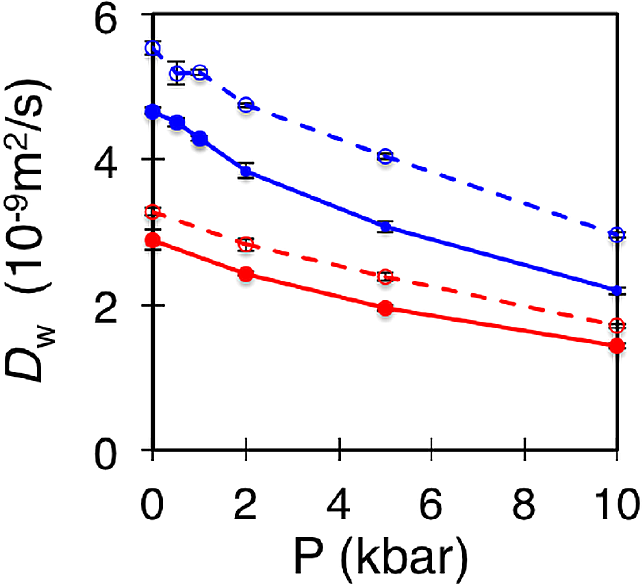}
}
\caption{(Color online) Diffusion coefficients of water as a function of pressure
in 24~ns molecular dynamics simulations with GB1 (filled,
connected by solid lines) and without GB1 (open, connected by dash
lines) at 0.15~M (blue) and 3~M (red) KCl. The error bars indicate
uncertainty amongst 4~ns blocks within each simulation.}
\label{Ichiye-fig08}
\end{wrapfigure}

So far, there have been a few studies of the effects of osmolytes
and kosmotropic/chaotropic salts on pressure-induced unfolding of
proteins, which indicate the effects consistent with the effects of small
solutes on unfolding by other means
\cite{Ichiye-ref79,Ichiye-ref80}. For instance, an FT-IR study of
staphylococcal nuclease indicates that kosmotropic salts protect
against pressure-induced unfolding in the volumetric contribution
while polyhydric alcohols and sugars stabilize against
pressure-induced unfolding in the energetic or entropic rather
than volumetric contribution, possibly due to preferential
hydration of the protein \cite{Ichiye-ref79}. In addition, the
24~ns molecular dynamics simulations of GB1 [Huang, Rodgers and Ichiye, unpublished results] were performed in 0.15~M and 3~M KCl,
to examine the effects of salt. Since the diffusion constant of
water, $D_{\text{w}}$, around GB1 decreases slightly with
pressure in 0.15~M KCl indicating an increase in solvent
viscosity, and decreases even more in 3~M KCl
(figure~\ref{Ichiye-fig08}), the effect of salt may also be in the
viscosity of the environment, although the magnitude of the
effects may be exaggerated due to the water model used in these
studies. This may protect the protein structure at higher pressures,
consistent with the observation that a halophile with high
intracellular salt concentration that normally lives at
atmospheric pressures is capable of surviving pressures up to at least
4~kbar \cite{Ichiye-ref28}. The latter suggests that further
studies might be necessary by examining the protein stability in
different salts and sugars at pressures beyond 8~kbar in order to
determine the pressure limits of microbial survival.

However, studies of the effects of osmolytes on the pressure
sensitivity of enzymatic activity have been even more rare.
Studies of co-solvents including glycerol, ethylene glycol,
sucrose, and methanol on \textit{M.~profunda} DHFR indicate that
the activity generally decreases with an increasing concentration, with
strong dependence on the dielectric constant and a weak dependence
on the viscosity \cite{Ichiye-ref73}, although the effects of
sucrose are weak. However, the decrease in the dielectric constant
is reflective not only of the decreased dielectric shielding but
also of the decreased long-range hydrogen-bonded order in the
liquid \cite{Ichiye-ref82} so care must be made in interpreting
these results. In addition, simple salts inhibit the activity of
\textit{E.~coli} DHFR \cite{Ichiye-ref83}. Altogether, these
studies suggest that although sugars and kosmotropic salts
increase the protein stability, they decrease the enzyme activity perhaps
by suppressing the protein flexibility by the effects such as an increased
viscosity, although further investigation is warranted.

Of natural piezolytes, it has been well demonstrated that the
levels of the compatible solute trimethylamine-N-oxide (TMAO) in
deep-sea metazoans correlate well with their depth of capture and
presumably their pressure tolerance \cite{Ichiye-ref84} and that
it stabilizes the proteins against pressure \cite{Ichiye-ref85}. The
effects of TMAO in stabilizing the proteins and in inducing the folding under
a variety of denaturing conditions is well established
\cite{Ichiye-ref86}. However, the effects of purported piezolytes
found in microbes on the enzyme activity, or even water, have not been
studied. Moreover, since the need to maintain flexibility over
stability appears important for enzyme activity, the protection
mechanisms of piezolytes might be different from sugars and
kosmotropic salts implicated for protein stability, or if they are
the same, would seem to indicate the necessity of extra
flexibility of the enzyme. The moderate piezophile \textit{P.~profundum}, which preferentially accumulates \textit{$\beta$}-hydroxybutarate under high pressure, was isolated at $\sim 260$
bar and can withstand pressures up to 900 bar, yet the activity of
its DHFR is reduced to $\sim 20${\%} of its atmospheric pressure
value \textit{in vitro} at 1000~bar~\cite{Ichiye-ref87}. This
suggests that other factors \textit{in vivo} such as a piezolyte
may enhance its activity at high pressures.

\subsection{Summary}

Microbes may use both genetic modifications of proteins as well as changes
in the intracellular milieu to protect against pressure. Genetic
modifications of proteins could involve changes in its material properties
or its volumetric properties. However, how changes in the intracellular
milieu affect the pressure effects on proteins and especially the enzyme activity is
much less understood, and could involve many possible factors such as
viscosity, water activity, changes in hydrogen bonding of water, and crowding.

\section{Conclusions}

The capability of microbes to function at a variety of extreme conditions, or
at least to preserve a sufficient structure to survive the extreme conditions, appears to
be a combination of the chemical composition of biological
macromolecules making up a microbe, the intracellular environment in which
they reside, and biochemical pathways for a repair of the damage due to the
extreme. Understanding the adaptations against pressure is still in its
infancy, but appears to involve more than adaptation by repair mechanisms.
In particular, adaptations for high-pressure environments that modify the
physical properties of the intracellular environment may play a significant
role by affecting the soft matter properties of the macromolecules
comprising the cell.

\section*{Acknowledgements}

QH, KNT, and TI are grateful for support from the National Science
Foundation through Grant \linebreak No.~CHE-1464766, the National Institutes
of Health through Grant No.~R21-GM104500, and from the McGowan
Foundation. JMR and RJH acknowledge support from the U.S.
Department of Energy/National Nuclear Security Administration
through Grant No.~DE-NA-0002006 for the Carnegie/DOE Alliance
Center (CDAC) and from the Alfred P.~Sloan Foundation through the
Deep Carbon Observatory. DHB is grateful to support from the
National Science Foundation (0801973, 0827051, and 1536776), the
National Aeronautics and Space Administration (NNX11AG10G), and
the Prince Albert~II Foundation (Project 1265). This work used
computer time on the Extreme Science and Engineering Discovery
Environment (XSEDE) granted via MCB990010, which is supported by
National Science Foundation Grant No.~OCI-1053575 and the Medusa
cluster, which is maintained by University Information Services at
Georgetown University. Anton computer time was provided by the
National Center for Multiscale Modelling of Biological Systems
(MMBioS) through grant P41GM103712-S1 from the National Institutes
of Health and the Pittsburgh Supercomputing Center (PSC). The
Anton machine at PSC was generously made available by D.E.~Shaw
Research.

\ukrainianpart

\title{Молекулярні перспективи для границь життя: \\ ензими під тиском}
\author{К. Гуанг\refaddr{label1}, К.Н. Трен\refaddr{label1}, Дж.М.~Роджерс\refaddr{label1,label2}, Д.Г. Бартлетт\refaddr{label3}, Р.Дж.~Гемлі\refaddr{label2}, Т.~Ічіє\refaddr{label1}}
\addresses{
\addr{label1} Хімічний факультет, Джорджтаунівський Університет, Вашінгтон, Округ Колумбія 20057, США
\addr{label2} Геофізична лабораторія, Інститут науки Карнегі, Вашінгтон, Округ Колумбія 20015-1305, США
\addr{label3} Інститут океанографії Скріппса, Каліфорнійський університет у Сан-Дієго, \\ Сан-Дієго, Каліфорнія 92093-0202, США
}

\makeukrtitle

\begin{abstract}
\tolerance=3000%
З чисто функціональної точки зору існування мікробів, що можуть рости в екстремальних
умовах, чи  ``екстремофілів'', приводить до питання як молекули, з яких створені ці мікроби, можуть підтримувати їх структуру і функцію. Тоді як мікроби, що живуть при екстремумах температури, були докладно досліджені, то ті, що живуть при екстремумах тиску, нехтувались, частково через труднощі в збиранні зразків та проведенні експериментів при нормальних умовах для мікроба. Однак, термодинамічні аргументи передбачають, що ефекти тиску могли б привести до різних організмових варіантів ніж ефекти температури. Очевидно, деякі з варіантів моли б бути серед властивостей конденсованої речовини у внутріклітинній рідині на додаток до генних
модифікацій макромолекул чи відновлювальних механізмів для макромолекул. В даному огляді
ефекти тиску на ензими, які є важливими протеїнам для росту і репродукції організму, та
деякі аргументи проти цих ефектів аналізуються та доповнюються результатами моделювання методом молекулярної динаміки. Метою огляду є закласти біологічну основу для досліджень цих систем під тиском з точки зору м'якої речовини.

\keywords ензими, гідростатичний тиск, внутріклітинне середовище

\end{abstract}

\begin{thebibliography}{99}

\bibitem{Ichiye-ref01} Winter~R., In: {Chemistry at Extreme Conditions},
 Manaa~M.R. (Ed.), 2005, Elsevier, Amsterdam, 29--82.

\bibitem{Ichiye-ref02} Meersman~F.,  Daniel~I.,  Bartlett~D.H.,  Winter~R.,  Hazael~R.,
McMillan~P.F., In: {Carbon in Earth}, Reviews in Mineralogical and Geochemistry Series Vol.~75,  Hazen~R.M.,  Jones~A.P.,  Baross~J.A. (Eds.),  Mineralogical Society of America,
Geochemical Society,  Chantilly, 2013, 607--648; \doi{10.2138/rmg.2013.75.19}.

\bibitem{Ichiye-ref03} Saiki R.,  Gelfand~D.,  Stoffel~S.,  Scharf~S.,  Higuchi~R.,  Horn~G., Mullis~K.,  Erlich~H., {Science}, 1988, \textbf{239}, 487; \doi{10.1126/science.2448875}.

\bibitem{Ichiye-ref04} Koga Y., {Archaea}, 2012, \textbf{2012}, 789652; \doi{10.1155/2012/789652}.

\bibitem{Ichiye-ref05} Russell N.J., {Philos. Trans. R. Soc. London, Ser. B}, 1990, \textbf{329}, 595; \doi{10.1098/rstb.1990.0034}.

\bibitem{Ichiye-ref06} Feller G.,  Gerday C., {Nat. Rev. Microbiol.}, 2003, \textbf{1},  200; \doi{10.1038/nrmicro773}.

\bibitem{Ichiye-ref07} Harding M.M.,  Ward L.G., Haymet A.D.J., {Eur. J. Biochem.}, 1999, \textbf{264}, 653; \doi{10.1046/j.1432-1327.1999.00617.x}.

\bibitem{Ichiye-ref08} Harding M.M.,  Anderberg P.I.,  Haymet A.D.J., {Eur. J. Biochem.}, 2003, \textbf{270}, 1381; \\ \doi{10.1046/j.1432-1033.2003.03488.x}.

\bibitem{Ichiye-ref09} Fang J.S.,  Zhang L.,  Bazylinski D.A., {Trends Microbiol.}, 2010, \textbf{18}, 413; \doi{10.1016/j.tim.2010.06.006}.

\bibitem{Ichiye-ref10} Prieur D.,  Jebbar M.,  Bartlett D.,  Kato C.,  Oger~P., In:
{Comparative High Pressure Biology},  Sebert~P. (Ed.),
CRC Press, Enfield, New Hampshire, 2009, 281--318.

\bibitem{Ichiye-ref11} Yayanos A.A., {Proc. Natl. Acad. Sci. U.S.A.}, 1986, \textbf{83}, 9542; \doi{10.1073/pnas.83.24.9542}.


\bibitem{Ichiye-ref12} Picard A.,  Daniel I., {Biophys. Chem.}, 2013, \textbf{183}, 30; \doi{10.1016/j.bpc.2013.06.019}.


\bibitem{Ichiye-ref13} Kallmeyer J.,  Pockalny R.,  Adhikari R.R.,  Smith D.C.,
 D'Hondt~S., {Proc. Natl. Acad. Sci. U.S.A.}, 2012, \textbf{109}, 16213; \doi{10.1073/pnas.1203849109}.


\bibitem{Ichiye-ref14} Whitman W.B.,  Coleman~D.C.,  Wiebe~W.J.,
{Proc. Natl. Acad. Sci. U.S.A.}, 1998, \textbf{95}, 6578; \\ \doi{10.1073/pnas.95.12.6578}.

\bibitem{Ichiye-ref15} Yayanos A.A., {Method. Microbiol.}, 2001, \textbf{30}, 615; \doi{10.1016/S0580-9517(01)30065-X}.

\bibitem{Ichiye-ref16} Kato C., In: {Extremophiles: Microbiology and Biotechnology},  Anitori~R.P. (Ed.), Caister Academic Press, Wymondham, 2012, Chapter~10, 233.

\bibitem{Ichiye-ref17} Yayanos A.A., {Annu. Rev. Microbiol.}, 1995, \textbf{49}, 777; \doi{10.1146/annurev.mi.49.100195.004021}.

\bibitem{Ichiye-ref18} Pal S., {Design of Arificial Human Joints and Organs},  Springer,
New York,  2014.

\bibitem{Ichiye-ref19} Kaye J.Z.,  Baross~J.A., {Appl. Environ. Microbiol.}, 2004,  \textbf{70}, 6220; \doi{10.1128/AEM.70.10.6220-6229.2004}.

\bibitem{Ichiye-ref20} Hay S.,  Evans R.M.,  Levy C.,  Loveridge E.J., Wang~X.,  Leys~D.,
Allemann~R.K.,  Scrutton~N.S., {ChemBioChem}, 2009, \textbf{10}, 2348; \doi{10.1002/cbic.200900367}.

\bibitem{Ichiye-ref21} Martin D.D.,  Bartlett~D.H.,  Roberts~M.F., {Extremophiles}, 2002, \textbf{6}, 507; \doi{10.1007/s00792-002-0288-1}.

\bibitem{Ichiye-ref22} Amrani A.,  Bergon A.,  Holota H.,  Tamburini C.,  Garel~M.,
Ollivier B.,  Imbert~J.,  Dolla~A.,  Pradel~N., {PLoS One}, 2014, \textbf{9}, e106831;
\doi{10.1371/journal.pone.0106831}.

\bibitem{Ichiye-ref23} Sharma A.,  Scott J.H.,  Cody G.D.,  Fogel M.L.,  Hazen R.M.,  Hemley R.J.,  Huntress~W.T., {Science}, 2002, \textbf{295}, 1514; \doi{10.1126/science.1068018}.

\bibitem{Ichiye-ref24} Yayanos A.A., {Science}, 2002, \textbf{297}, 295; \doi{10.1126/science.297.5580.295a}.

\bibitem{Ichiye-ref25} Vanlint D.,  Mitchell R.,  Bailey E.,  Meersman F.,  McMillan~P.F.,
Michiels C.W., Aertsen~A., {mBio}, 2011, \textbf{2}, e00130-10; \doi{10.1128/mBio.00130-10}.

\bibitem{Ichiye-ref26} Hazael R.,  Foglia F.,  Kardzhaliyska L.,  Daniel I.,  Meersmen~F.,
McMillan P., {Front. Microbiol.}, 2014, \textbf{5}, 612; \doi{10.3389/fmicb.2014.00612}.

\bibitem{Ichiye-ref27} Molina-H{\"o}ppner~A.,  Doster W., Vogel~R.F.,  G{\"a}nzle~M.G., {Appl. Environ. Microbiol.}, 2004, \textbf{70}, 2013; \\ \doi{10.1128/AEM.70.4.2013-2020.2004}.

\bibitem{Ichiye-ref28} Kish~A.,  Griffin~P.L., Rogers~K.L., Fogel~M.L.,  Hemley~R.J.,
Steele~A., {Extremophiles}, 2012, \textbf{16}, 355; \\ \doi{10.1007/s00792-011-0418-8}.

\bibitem{Ichiye-ref29} Valenti P.,  Bodnar R.J.,  Schmidt C., {Geochim. Cosmochim. Acta}, 2012, \textbf{92}, 117; \doi{10.1016/j.gca.2012.06.007}.

\bibitem{Ichiye-ref30} Yoshimura~Y.,  Mao H.-k.,  Hemley~R.J., {Chem. Phys. Lett.}, 2004, \textbf{400}, 511; \doi{10.1016/j.cplett.2004.10.139}.

\bibitem{Ichiye-ref31} Abramson E.H., {Phys. Rev. E}, 2007, \textbf{76}, 051203; \doi{10.1103/PhysRevE.76.051203}.

\bibitem{Ichiye-ref32} Kestin J.,  Khalifa H.E.,  Abe Y.,  Grimes C.E.,  Sookiazian~H.,
Wakeham W.A., {J. Chem. Eng. Data}, 1978, \textbf{10}, 328; \doi{10.1021/je60079a011}.

\bibitem{Ichiye-ref33} Kestin J.,  Khalifa H.E.,  Correia R.J., {J. Phys. Chem. Ref. Data}, 1981, \textbf{10}, 57; \doi{10.1063/1.555640}.

\bibitem{Ichiye-ref34} Yayanos A.A., Dietz A.S.,  Van Boxtel R., {Proc. Natl. Acad. Sci. U.S.A.}, 1981, \textbf{78}, 5212; \doi{10.1073/pnas.78.8.5212}.

\bibitem{Ichiye-ref35} Hauben K.J.,  Bartlett D.H.,  Soontjens C.C.,  Cornelis K.,  Wuytack~E.Y.,  Michiels~C.W., {Appl. Environ. Microbiol.}, 1997, \textbf{63}, 945.

\bibitem{Ichiye-ref36} Jofr\'e A.,  Aymerich T.,  Bover-Cid S.,
     Garriga M., {Int. Microbiol.}, 2010, \textbf{13},  105.

\bibitem{Ichiye-ref50} Best R.B.,  Zhu~X.,  Shim J.,  Lopes P.,  Mittal J.,  Feig~M.,
MacKerell~A.D.~Jr., {J. Chem. Theory Comput.}, 2012, \textbf{8}, 3257; \doi{10.1021/ct300400x}.

\bibitem{Ichiye-ref49} MacKerell A.D. Jr.,  Bashford D., Bellot M.,  Dunbrack R.L. Jr.,
Field M.J.,  Fischer~S.,  Gao~J.,  Guo~H.,  Ha~S.,  Joseph~D.,  Kuchnir~K.,
 Kuczera~K.,  Lau~F.T.K.,  Mattos~M.,  Michnick~S.,  Nguyen~D.T., Ngo~T.,  Prodhom~B., Roux~B.,  Schlenkrich~M.,  Smith~J.,  Stote~R., Straub~J.,  Wiorkiewicz-Kuczera~J.,
 Karplus~M., {J. Phys. Chem. B}, 1998, \textbf{102}, 3586; \doi{10.1021/jp973084f}.


\bibitem{Ichiye-ref81} Jorgensen W.L.,  Chandrasekhar~J.,  Madura~J.D.,  Impey~R.W.,
Klein~M.L., {J. Chem. Phys.}, 1983, \textbf{79}, 926; \\ \doi{10.1063/1.445869}.

\bibitem{Ichiye-ref51} Horn H.W.,  Swope W.C.,  Pitera J.W.,  Madura J.D.,  Dick~T.J.,  Hura G.L.,  Head-Gordon~T., {J. Chem. Phys.}, 2004, \textbf{120}, 9665; \doi{10.1063/1.1683075}.

\bibitem{Ichiye-ref37} Bridgman P.W., {J. Biol. Chem.}, 1914, \textbf{19}, 511.


\bibitem{Ichiye-ref39} Roche J.,  Caro J.A.,  Noberto D.R.,  Barthe~P.,  Roumestand~C.,
Schlessman~J.L.,  Garcia~A.E.,  Garcia-Moreno~B.E.,  Royer~C.A., {Proc. Natl. Acad. Sci. U.S.A.}, 2012, \textbf{109}, 6945; \doi{10.1073/pnas.1200915109}.

\bibitem{Ichiye-ref38} Frye K.J.,  Royer C.A., {Protein Sci.}, 1998, \textbf{7}, 2217; \doi{10.1002/pro.5560071020}.

\bibitem{Ichiye-ref40} Collins M.D.,  Hummer G.,  Quillin~M.L.,  Matthews~B.W.,  Gruner~S.M., {Proc. Natl. Acad. Sci. U.S.A.}, 2005, \textbf{46}, 16668; \doi{10.1073/pnas.0508224102}.

\bibitem{Ichiye-ref41} Ando N.,  Barstow B.,  Baase~W.A.,  Fields~A.,  Matthews~B.W.,
Gruner S.M., {Biochem.}, 2008, \textbf{47}, 11097; \\ \doi{10.1021/bi801287m}.

\bibitem{Ichiye-ref42} Nucci N.V.,  Fuglestad~B.,  Athanasoula~E.A.,  Wand~J.A.,
{Proc. Natl. Acad. Sci. U.S.A.}, 2014, \textbf{111}, 13846; \\ \doi{10.1073/pnas.1410655111}.

\bibitem{Ichiye-ref43} Panick G.,  Malessa R.,  Winter R.,  Rapp G.,  Frye K.J.,  Royer~C.A., {J. Mol. Biol.}, 1998, \textbf{275}, 389; \\ \doi{10.1006/jmbi.1997.1454}.

\bibitem{Ichiye-ref44} Hummer G.,  Garde S.,  Garcia A.E.,  Paulaitis M.E., Pratt~L.R.,
{Proc. Natl. Acad. Sci. U.S.A.}, 1998, \textbf{95}, 1552; \doi{10.1073/pnas.95.4.1552}.

\bibitem{Ichiye-ref45} Boonyaratanakornkit B.B.,  Park C.B.,  Clark D.S., {Biochim. Biophys. Acta, Protein Struct. Mol. Enzymol.}, 2002, \textbf{1595}, 235; \doi{10.1016/S0167-4838(01)00347-8}.

\bibitem{Ichiye-ref46} Nagae T.,  Kawamura T.,  Chavas L.M.G.,  Niwa K.,  Hasegawa M.,  Kato~C.,  Watanabe N., {Acta Crystallogr. D}, 2012, \textbf{68}, 300; \doi{10.1107/S0907444912001862}.

\bibitem{Ichiye-ref47} Ohmae E.,  Tatsuka M.,  Abe F.,  Kato C.,  Tanaka N.,  Kunugi~S.,
Gekko~K., {Biochim. Biophys. Acta, Proteins Proteomics}, 2008, \textbf{1784}, 1115; \doi{10.1016/j.bbapap.2008.04.005}.

\bibitem{Ichiye-ref48} Wilton D.J.,  Tunnicliffe R.B.,  Kamatari Y.O.,  Akasaka K.,
Williamson M.P., {Proteins Struct. Funct. Bioinf.}, 2008,
\textbf{71}, 1432; \doi{10.1002/prot.21832}.

\bibitem{Ichiye-ref52} Kundrot C.E.,  Richards F.M., {J. Mol. Biol.}, 1987, \textbf{193}, 157; \doi{10.1016/0022-2836(87)90634-6}.

\bibitem{Ichiye-ref53} Ascone I.,  Savino C.,  Kahn R.,  Fourme R., {Acta Crystallogr. D}, 2010, \textbf{66}, 654; \doi{10.1107/S0907444910012321}.


\bibitem{Ichiye-ref54} Meinhold L.,   Smith J.C., {Phys. Rev. E}, 2005, \textbf{72},  061908; \doi{10.1103/PhysRevE.72.061908}.

\bibitem{Ichiye-ref55} Meinhold L.,  Smith J.C.,  Kitao A.,  Zewail A.H.,
{Proc. Natl. Acad. Sci. U.S.A.}, 2005, \textbf{104}, 17261; \\ \doi{10.1073/pnas.0708199104}.

\bibitem{Ichiye-ref56} Barstow B.,  Ando N.,  Kim C.U.,  Gruner S.M.,
{Proc. Natl. Acad. Sci. U.S.A.}, 2008, \textbf{105}, 13362; \\ \doi{10.1073/pnas.0802252105}.

\bibitem{Ichiye-ref57} Evans R.M.,  Behiry E.M.,  Tey L.-H.,  Guo J.,  Loveridge E.J.,
Allemann R.K., {ChemBioChem}, 2010, \textbf{11}, 2010; \doi{10.1002/cbic.201000341}.

\bibitem{Ichiye-ref58} Bae E.,  Phillips G.N.J., {Proc. Natl. Acad. Sci. U.S.A.}, 2004, \textbf{103}, 2132; \doi{10.1073/pnas.0507527103}.

\bibitem{Ichiye-ref59} Ruan Q.,  Ruan K.,  Balny C.,  Glaser M.,  Mantulin W.W., {Biochem.}, 2001, \textbf{40}, 14706; \doi{10.1021/bi010308i}.

\bibitem{Ichiye-ref60} Sawaya M.R.,  Kraut J., {Biochem.}, 1997, \textbf{36}, 586; \doi{10.1021/bi962337c}.

\bibitem{Ichiye-ref61} Osborne M.J.,  Schnell J.,  Benkovic S.J.,  Dyson H.J.,  Wright P.E., {Biochem.}, 2001, \textbf{40}, 9846; \doi{10.1021/bi010621k}.

\bibitem{Ichiye-ref62} Kitahara R.,  Sareth S., Yamada H.,  Ohmae E., Gekko K.,  Akasaka~K., {Biochem.}, 2000, \textbf{39}, 12789; \\ \doi{10.1021/bi0009993}.

\bibitem{Ichiye-ref63} Kitahara R.,  Yokoyamaa S.,  Akasaka K., {J. Mol. Biol.}, 2005,  \textbf{347}, 277; \doi{10.1016/j.jmb.2005.01.052}.

\bibitem{Ichiye-ref64} Kasahara R.,  Sato T.,  Tamegai H.,  Kato C., {Biosci. Biotechnol., Biochem.}, 2009, \textbf{73}, 2541; \doi{10.1271/bbb.90448}.

\bibitem{Ichiye-ref65} Nagae T.,  Kato C.,  Watanabe N., {Acta Crystallogr. F}, 2012, \textbf{68}, 265; \doi{10.1107/S1744309112001443}.

\bibitem{Ichiye-ref66} Bae E.,  Phillips G.N.J., {J. Biol. Chem.}, 2004, \textbf{279}, 28202; \doi{10.1074/jbc.M401865200}.

\bibitem{Ichiye-ref67} Berezovsky I.N.,  Shakhnovich E.I., {Proc. Natl. Acad. Sci. U.S.A.}, 2005, \textbf{102}, 12742; \doi{10.1073/pnas.0503890102}.

\bibitem{Ichiye-ref68} Georlette D.,  Blaise V.,  Collins T.,  D'Amico S.,  Gratia E.,  Hoyoux~A.,  Marx~J.-C.,  Sonan~G.,  Feller~G.,  Gerday~C., {FEMS
Microbiol. Rev.}, 2004, \textbf{28}, 25; \doi{10.1016/j.femsre.2003.07.003}.

\bibitem{Ichiye-ref69} Meersman F.,  Smeller L.,  Heremans K., {Biophys. J.}, 2002, \textbf{82}, 2635; \doi{10.1016/S0006-3495(02)75605-1}.

\bibitem{Ichiye-ref70} Gross M.,  Jaenicke R., {Eur. J. Biochem.}, 1994, \textbf{221},  617; \doi{10.1111/j.1432-1033.1994.tb18774.x}.

\bibitem{Ichiye-ref71} Kato~C.,  Li L.,  Nogi Y.,  Nakamura Y.,  Tamaoka~J.,  Horikoshi~K.,
{Appl. Environ. Microbiol.}, 1998, \textbf{64}, 1510.

\bibitem{Ichiye-ref72} Abboud R.,  Popa R.,  Souza-Egipsy V.,  Giometti C.S.,
Tollaksen~S., Mosher~J.J.,  Findlay~R.H.,  Nealson~K.H., {Appl. Environ.
Microbiol.}, 2005, \textbf{71}, 811; \doi{10.1128/AEM.71.2.811-816.2005}.


\bibitem{Ichiye-ref74} Ohmae E.,  Murakami~C.,  Tate~S.-i.,  Gekko~K.,  Hata~K.,  Akasaka~K.,  Kato~C., {Biochim. Biophys. Acta, Proteins Proteomics}, 2012, \textbf{1824}, 511; \doi{10.1016/j.bbapap.2012.01.001}.

\bibitem{Ichiye-ref73} Loveridge E.J.,  Tey L.-H.,  Behiry~E.M.,
     Dawson~W.H.,  Evans~R.M.,  Whittaker~S.B.-M.,  G{\"u}nther~U.L.,
    Williams~C., Crump~M.P.,  Allemann~R.K., {J. Am. Chem.
    Soc.}, 2011, \textbf{113}, 20561; \doi{10.1021/ja208844j}.

\bibitem{Ichiye-ref75} Murakami C., Ohmae E.,  Tate S.-i.,  Gekko K.,  Nakasone~K.,
Kato~C., {Extremophiles}, 2011, \textbf{15}, 165; \\ \doi{10.1007/s00792-010-0345-0}.

\bibitem{Ichiye-ref76} Bartlett D.H., {Biochim. Biophys. Acta, Protein Struct. Mol. Enzymol.}, 2002, \textbf{1595}, 367; \\ \doi{10.1016/S0167-4838(01)00357-0}.

\bibitem{Ichiye-ref77} Alazard D.,  Dukan~S.,  Urios~A.,
    Verh{\'e}~F.,  Bouabida~N.,  Morel~F.,  Thomas~P.,  Garcia~J.-L.,
     Ollivier~B., {Int. J. Syst. Evol. Microbiol.}, 2003,     \textbf{53}, 173; \doi{10.1099/ijs.0.02323-0}.

\bibitem{Ichiye-ref78} Engel M.B.,  Catchpole~H.R., {Cell Biol. Int.}, 2005, \textbf{29}, 616; \doi{10.1016/j.cellbi.2005.03.024}.

\bibitem{Ichiye-ref79} Herberhold H.,  Royer~C.A., Winter~R., {Biochem.}, 2004, \textbf{43}, 3336; \doi{10.1021/bi036106z}.

\bibitem{Ichiye-ref80} Krywka C.,  Sternemann C.,  Paulus M.,  Tolan M.,  Royer~C.A.,
Winter~R., {ChemPhysChem}, 2008, \textbf{9}, 2809; \\ \doi{10.1002/cphc.200800522}.


\bibitem{Ichiye-ref82} Tan M.-L., Cendagorta~J.R.,  Ichiye~T., {J. Chem. Phys.}, 2014, \textbf{141}, 244504; \doi{10.1063/1.4904263}.

\bibitem{Ichiye-ref83} Ohmae E.,  Miyashita Y.,  Tate~S.-i.,  Gekko~K.,  Kitazawa~S.,  Kitahara~R.,  Kuwajima~K., {Biochim. Biophys. Acta, Proteins Proteomics}, 2013, \textbf{1834}, 511; \doi{10.1016/j.bbapap.2013.09.024}.

\bibitem{Ichiye-ref84} Yancey P.H.,  Gerringer~M.,  Rowden~A.A.,  Drazen~J.C.,  Jamieson~A., {Proc. Natl. Acad. Sci. U.S.A.}, 2014, \textbf{111}, 4461; \doi{10.1073/pnas.1322003111}.

\bibitem{Ichiye-ref85} Yancey P.H.,  Fyfe-Johnson~A.L., Kelly~R.H.,  Walker~V.P.,
 Au{\~n}{\'o}n~M.T., {J. Exp. Zool.}, 2001, \textbf{289}, 172; \\ \doi{10.1002/1097-010X(20010215)289:3<172::AID-JEZ3>3.0.CO;2-J}.

\bibitem{Ichiye-ref86} Bolen D.W.,  Rose~G.D., {Annu. Rev. Biochem.}, 2008, \textbf{77}, 339; \doi{10.1146/annurev.biochem.77.061306.131357}.

\bibitem{Ichiye-ref87} Murakami C.,  Ohmae E.,  Tate~S.-i.,  Gekko~K.,  Nakasone~K.,  Kato~C., {J. Biochem.}, 2010, \textbf{147}, 591; \\ \doi{10.1093/jb/mvp206}.

\end{thebibliography}
\end{document}